\documentclass{article}

\usepackage{arxiv}

\usepackage[utf8]{inputenc} 
\usepackage[T1]{fontenc}    
\usepackage{hyperref}       
\usepackage{url}            
\usepackage{booktabs}       
\usepackage{amsfonts}       
\usepackage{nicefrac}       
\usepackage{microtype}      
\usepackage{lipsum}
\usepackage{amsmath, amssymb}
\usepackage{comment}
\usepackage{multirow}
\usepackage{graphicx} 

\title{TPLVM: Portfolio Construction by Student's $t$-process Latent Variable Model}

\author{
  Yusuke Uchiyama\\
  MAZIN Inc.,\\
  3-29-14 Nishi-Asakusa, Tito city, Tokyo 111-0035 Japan \\
  \texttt{uchiyama@mazin.tech}
   \And
  Kei Nakagawa\\
  Innovation Lab\\
  Nomura Asset Management Co., Ltd.\\
  1-11-1 Nihonbashi, Chuo-ku, Tokyo, 103-8260, Japan \\
  \texttt{kei.nak.0315@gmail.com} 
}

\begin{document}
\maketitle

\begin{abstract}
Optimal asset allocation is a key topic in modern finance theory.  To realize the optimal asset allocation on investor's risk aversion, various portfolio construction methods have been proposed.  Recently, the applications of machine learning are rapidly growing in the area of finance.  In this article, we propose the Student's $t$-process latent variable model (TPLVM) to describe non-Gaussian fluctuations of financial timeseries by lower dimensional latent variables.  Subsequently, we apply the TPLVM to minimum-variance portfolio as an alternative of existing nonlinear factor models.  To test the performance of the proposed portfolio, we construct minimum-variance portfolios of global stock market indices based on the TPLVM or Gaussian process latent variable model.  By comparing these portfolios, we confirm the proposed portfolio outperforms that of the existing Gaussian process latent variable model.
\end{abstract}


\section{Introduction}
\label{sec:intro}
%
Estimation of covariance matrix of timeseries plays a dominant role in applications of modern financial theory.  The optimization of mean-variance portfolio, which is one of the pioneering works of the modern finance theory~\cite{markowitz1952portfolio}, is based on the covariance matrix of the multi-dimensional timeseries of return of assets.  Since the return of assets are modelled by non-stationary stochastic processes, the covariance matrix should be estimated as a time-dependent symmetric matrix.  In practice, we often estimate the covariance matrix by empirical time averaging, because of the lack of complete information of the corresponding probabilistic space.  It is however pointed out that time averaging often causes serious estimation error of the covariance matrix in the case of larger assets~\cite{nakagawa2018risk,engle2019large}.  To overcome this problem, several inference methods are proposed from the point of view of the random matrix theory~\cite{ledoit2012nonlinear,ledoit2017nonlinear}.  \par
%
With the aid of recently growing machine learning techniques, we can improve the accuracy of the estimation of the covariance matrix~\cite{chen2017covmat,wu2013covmat}.  Furthermore, the applications of the machine learning techniques have been spreading in both theoretical and practical financial problems~\cite{atsalakis2009surveying,cavalcante2016computational}.  The prediction of the future price is implemented by the deep neural networks of various modeling~\cite{nakagawa2018deep,nakagawa2019deep}.  The Gaussian process is used as a model of dynamics of the covariance matrix of multi-dimensional timeseries.  In the literature of option pricing theory, the model of the volatility of a risky asset is given by the Gaussian process~\cite{wu2014gpvm}.  In particular, the application of the machine learning techniques for the portfolio optimization has attracted the interest of both academia and industry~\cite{shen2015folio,song2017folio}.  \par
In the field of mathematical finance, stochastic volatility models have been utilized in estimating dynamic covariance matrix of the return of assets.  One of the most popular conditional volatility models is the generalized autoregressive conditional heteroscedasticity (GARCH) model~\cite{bollerslev1986generalized}, which describes the volatility clustering of the return of assets.  To introduce a time-varying correlation structure to these conditional volatility models, the dynamic conditional correlation (DCC) GARCH model has been proposed~\cite{engle2002dynamic}.  The parameters of the GARCH and DCC GARCH can be estimated by the method of maximum likelihood.    \par
On the other hand, in the literature of the machine learning, some kinds of latent variable models can be utilized to infer the dynamics of the covariance matrix.  Recently, the Gaussian process latent variable model (GPLVM) has been employed to the problem of the portfolio optimization, where latent variables are introduced as factors of return of the assets.  Namely, this model can be interpreted as a latent variable factor model~\cite{nirwan2019applications}.  \par
Despite these practical applications, we should reconsider the assumption and validation of the use of the GPLVM for finance because the GPLVM assumes that observed data follows the Gaussian distribution.  In the most case of financial problems, the return of assets is regarded as an observed variable.  It is well known that the fluctuations of the return of assets follow non-Gaussian distributions\cite{mandelbrot1997variation}.  To describe such fluctuations, some fat-tailed distributions have been presented and applied to the financial timeseries.  Thus, the GPLVM should be extended to fat-tailed distributions when we use it for the financial problems.  \par
In this paper, we propose Student's-$t$ process latent variable model (TPLVM) as an extension of the GPLVM.  This model is developed based on the Student's $t$-distribution, which is a symmetric fat-tailed distribution.  Since the Student's $t$-distribution converges to the Gaussian distribution with the limit of a parameter, degree of freedom, the TPLVM includes the GPLVM as a special case.  To use the TPLVM in practice, as well as the GPLVM, we derive its predictive distribution as closed form and an estimator of hyper parameters by the variational inference in Bayesian sense.  \par
The reminder of this paper is organized as follows.  Chapter~\ref{sec:GP} gives a brief introduction the GPLVM including the Gaussian process with the concept of kernel functions.  In Chap. 3, we introduce the formula of TPLVM, which consists of the kernel functions, predictive distribution and variational inference for estimating hyper parameters.  As a preliminary preparation of finance, we explain the basis of factor model and portfolio optimization in Chap. 4.  Chapter 5 implements portfolio optimization, where we compare the performance of the GPLVM and TPLVM.  Chapter 6 is dedicated to conclusions and future works.
\section{Short review of Gaussian process}
\label{sec:GP}
\subsection{Gaussian process}
The Gaussian process, a kind of stochastic processes, is a non-parametric method of machine learning method~\cite{rasmussen2003gaussian,williams2006gaussian}.  This has been firstly introduced to describe random dynamics such as a fluctuating pollen on water surface known as Brownian motion~\cite{einstein1905brownianmotion}.  Without loss of generality, the argument of the Gaussian process can be extended from one-dimensional time to multi-dimensional feature space.  In this chapter, we provide a short review of the Gaussian process for multi-dimensional features as the preliminary preparation of the proposed model.  \par
For a sequence of input features $\{x_1, x_2, {\cdots}, x_n\}$, a stochastic process $f({\cdot})$ is the Gaussian process when the sequence of random variables $\{f(x_1), f(x_2), {\cdots}, f(x_n)\}$ is sampled from the multivariate Gaussian distribution.  In general, the form of the multivariate Gaussian distribution is determined by the mean vector and covariance matrix.  Likewise, the Gaussian process are specified by the mean and covariance function of input features.  Thus, the Gaussian process is regarded as the representation of the infinite dimensional Gaussian distribution.  \par
The mean and covariance functions are defined as follows:
\begin{align}
&m(x)=\mathbb{E}[f(x)],
\label{eq:MEANFUNC} \\
&k(x,x')=\mathbb{E}[(f(x)-m(x))(f(x')-m(x'))],
\label{COVFUNC}
\end{align}
where the operator $\mathbb{E}[{\cdot}]$ denotes expectation, $m({\cdot})$ and $k({\cdot},{\cdot})$ are respective mean and covariance functions.  The mean vector and covariance matrix of the Gaussian process for given dataset are represented by 
\begin{align}
&m_i = m(x_i), \\
&K_{i,j}=k(x_i,x_j).
\end{align}
On these setting, the stochastic process $f({\cdot})$ is sampled from $\mathcal{N}(m({\cdot}),K({\cdot}, {\cdot}))$.  In this situation, the stochastic process $f({\cdot})$ is the Gaussian process expressed as $f{\sim}\mathcal{GP}(m,K)$.  The covariance function satisfies to be symmetric and positive definite, and thus is also called as kernel function.  In the literature of the Gaussian process, the covariance matrix is often called as kernel matrix.  The mathematical characteristics of the kernel functions are explained in~\cite{hofmann2008kernel}.  \par
Given an additional dataset $\mathcal{D}^*=\{x_1', x_2', {\cdots}, x_{n'}'\}$, the corresponding outputs $\{y_1', y_2', {\cdots}, y_{n'}'\}$ can be predicted by the conditional Gaussian process with prior dataset $\mathcal{D}=\{(x_1,y_1), (x_2,y_2), {\cdots}, (x_n,y_n)\}$.  With notations that $X=[x_1, x_2, {\cdots}, x_n]^T$, $X^*=[x_1', x_2', {\cdots}, x_{n'}]^T$ and $Y=[y_1,y_2,{cdots},y_n]$, the predictive distribution of the conditional Gaussian process is also given by the Gaussian process $\mathcal{GP}(f^*, K^*)$, where
\begin{align}
&f^*=m_X+K_{X^*,X}K_{X,X}^{-1}Y,
\label{eq:PEDMEAN} \\
&K^*=K_{X^*,X^*}-K_{X^*,X}K_{X,X}^{-1}K_{X,X^*}.  
\label{eq:PRECOV}
\end{align}
In Eqs.~(\ref{eq:PEDMEAN}) and (\ref{eq:PRECOV}), it is seen that the covariance functions propagate the information about $\mathcal{D}$ to $\mathcal{D}^*$.  Hence, the covariance functions play the dominant role in the use of the Gaussian process.  \par
\subsection{Gaussian process latent variable model}
In the literature of big data analysis, it is often expected that observed variables can be explained by lower dimensional latent variables.  For this purpose, various methods of dimension reduction have been developed.  One of the most popular methods is the principal component analysis (PCA), which extracts latent variables by the singular value decomposition.  To extend the PCA for nonlinear and random data, the Gaussian process latent variable model (GPLVM) has been proposed~\cite{lawrence2004gaussian}.  The GPLVM expresses the nonlinearity of both observed and latent variables by the covariance function.  The randomness is assumed to be originate from the Gaussian distribution.  \par
To describe an observed variable $y{\in}\mathbb{R}^D$, we introduce a latent variable $x{\in}\mathbb{R}^Q$ with $Q<D$, and a nonliner map $f:\mathbb{R}^Q{\to}\mathbb{R}^D$ with a $Q$-dimensional noise ${\epsilon}{\sim}\mathcal{N}(0, {\sigma}_0I)$ as
\begin{equation}
y=f(x)+{\varepsilon}. 
\label{eq:LTV}
\end{equation}
For this latent variable model, we assume that the nonlinear map $f({\cdot})$ is sampled from the Gaussian process as $f{\sim}\mathcal{N}(0, K)$.  This model is known as the GPLVM.  For the sake of brevity, we introduce notations for the set of latent and observed variables as $X=[x_1,x_2,{\cdots},x_N]^T$ and $Y=[y_1,y_2,{\cdots},y_N]^T$.  Assume that the columns of the observed matrix $Y{\in}\mathbb{R}^{N{\times}D}$ are samples from the independently identical distributed Gaussian distributions which have the covariance functions with respect to the latent variable matrix $X{\in}\mathbb{R}^{N{\times}Q}$, the probability density function of the GPLVM is introduced as follows:
\begin{equation}
p(Y|X)=\frac{1}{(2{\pi})^{ND/2}|K_{X,X}|^{D/2}}\exp{\left(-\frac{1}{2}Y^TK_{X,X}^{-1}Y\right)}.
\label{eq:LKLTV}
\end{equation}
In the GPLVM, hyperparameters of the covariance functions and latent variables are inferred by several existing methods such as gradient methods, variational inference and Markov Chain Monte Carlo methods. 
\section{Proposed model: Student's $t$-process latent variable model}
\label{sec:TPLVM}
\subsection{Introduction of the Student's $t$-process}
The Gaussian process has diverse applications in the fields of computer science, robotics and others.  However, it seems not to be applicable to problems in finance because the fluctuations of the financial data follow non-Gaussian distributions with fat-tails.  It is thus necessary to extend the methods of the Gaussian process non-Gaussian stochastic processes with fat-tails.  \par
For this purpose, the Student's $t$-process was proposed as a generalization of the Gaussian process~\cite{shah2014student}.  This stochastic process follows the Student's $t$-distribution, of which tails show power-law behaviours.  As with the Gaussian process, the Student's $t$-process is specified by the mean and covariance functions.  Given the mean and covariance functions, the probability density function of the Student's $t$-process is defined as
\begin{equation}
\mathcal{T}(m, K, {\nu})=\frac{{\Gamma}\left(\frac{{\nu}+N}{2}\right)}{[({\nu}-2){\pi}]^{\frac{N}{2}}{\Gamma}\left(\frac{\nu}{2}\right)|K|^{\frac{1}{2}}}\left[1+\frac{1}{{\nu}-2}(y-m)^TK^{-1}(y-m)\right]^{-\frac{{\nu}+N}{2}},
\label{eq:TDIST}
\end{equation}
where ${\Gamma}({\cdot})$ is the multivariate gamma function and the positive real parameter ${\nu}$ is degrees of freedom.  In this setting, the stochastic process $f({\cdot})$ is the Student's $t$-process expressed as $f{\sim}\mathcal{TP}(m,K;{\nu})$.  Note that the Student's $t$-process converges to the Gaussian process at the limit of ${\nu}{\to}{\infty}$.  \par
The conditional distribution of the Student's $t$-process can be also derived analytically and given as the conditional Student's $t$-distribution.  Namely, we can update the mean and covariance functions and the degrees of freedom from the conditional distribution.  Through cumbersome calculations, the renewal formulas of the mean and covariance functions and the degrees of freedom are derived as follows:
\begin{align}
&m^*=m + K_{X^*,X}K_{X,X}^{-1}Y,
\label{eq:TPREDMU} \\
&K^*=\frac{{\nu}-{\beta}-2}{{\nu}-N-2}\left[K_{X^*,X^*}-K_{X^*,X}K_{X,X}^{-1}K_{X,X^*}\right],
\label{eq:TPREDK} \\
&{\beta}=(Y-m_X)^TK_{X,X}^{-1}(Y-m_X),
\label{eq:TPBETA} \\
&{\nu}^*={\nu}+N.
\label{eq:TPREDNU}
\end{align}
It is seen that the renewal formula of the covariance function explicitly depends on the number of observed variables, which property does not appear in the case of the Gaussian process.  Hence, the Student's $t$-process is regarded to utilize prior information more effectively than the Gaussian process.  
\subsection{Student's $t$-process latent variable model}
To extend the GPLVM to stochastic processes following non-Gaussian distributions, we propose the Student's-$t$ process latent variable model (TPLVM).  Suppose an observed variable $y{\in}\mathbb{R}^D$ is explained by a low dimensional latent variable $x{\in}\mathbb{R}^Q\;(Q<D)$ by a nonlinear map $f:\mathbb{R}^D{\to}\mathbb{R}^Q,\;f{\sim}\mathcal{TP}(m,K;{\nu})$, the TPLVM is introduced as follows:
\begin{equation}
p(Y|X)=\frac{{\Gamma}\left(\frac{{\nu}+D}{2}\right)}{[({\nu}-2){\pi}]^{\frac{D}{2}}{\Gamma}\left(\frac{\nu}{2}\right)|K_{X,X}|^{\frac{1}{2}}}\left[1+\frac{1}{{\nu}-2}(Y-m_X)^TK_{X,X}^{-1}(Y-m_X)\right]^{-\frac{{\nu}+D}{2}}.
\label{eq:TPLTV}
\end{equation}
The nonlinear dependency of the latent matrix $X{\in}\mathbb{R}^{N{\times}Q}$ is given through the covariance matrix.  It is expected that the TPLVM provides a robust estimation especially for observed data with large fluctuations because the Student's $t$-distribution can capture large deviated data from the Gaussian distribution in its sampling.  \par
As with the GPLVM, the latent variable and hyperparameters of the TPLVM can be estimated from its likelihood.  The logarithmic likelihood of the TPLVM is given as
\begin{align}
\log{p(Y|X)}&=\log{{\Gamma}\left(\frac{{\nu}+D}{2}\right)}-\frac{D}{2}\log{[({\nu}-2){\pi}]}
-\log{{\Gamma}\left(\frac{\nu}{2}\right)}-\frac{1}{2}\log{|K_{X,X}|} \nonumber \\
&-\frac{{\nu}+D}{2}\log{\left[1+ \frac{1}{{\nu}-2}(Y-m_X)^TK_{X,X}^{-1}(Y-m_X)\right]},
\label{eq:LKHDTPLTV}
\end{align}
By means of existing optimization methods, we can estimate the latent variables and hyperparameters of the covariance function and the degrees of freedom.  However, it is known that the optimization of the covariance function with respect to the latent variables often induces numerical instability because of its complexity.  Hence, we should carefully select the initial values of optimization procedures and repeat with diverse seeds of the initial values to refuse dropping in local minima.
\subsection{Variational inference}
To overcome the shortcomings of the method of maximum-likelihood, we utilize the method of variational inference~\cite{damianou2016variational}.  Instead of optimizing the logarithmic likelihood in Eq.~(\ref{eq:LKHDTPLTV}), we consider that of posterior $p(X|Y)=p(Y|X)p(X)/p(Y)$ in the Bayesian sense.  In solving the optimization problem with respect to the posterior, we try to approximate $p(X|Y)$ by $q(X)$.  As a measure of the difference between two probability density functions, we introduce the Kullback-Leibler (KL) divergence as follows:
\begin{equation}
\mathrm{KL}[q(X)||p(X|Y)]={\int}\log{\frac{q(X)}{p(X|Y)}}q(X)\mathrm{d}X.
\label{eq:KLDIV-1}
\end{equation}
With the use of the Bayes theorem, the KL divergence is alternatively represented as
\begin{equation}
\mathrm{KL}[q(X)||p(X|Y)]=-{\int}\log{\frac{p(Y|X)p(X)}{q(X)}}q(X)\mathrm{d}X+\log{p(Y)}.
\label{eq:KLDIV-2}
\end{equation}
Since the second term in the right hand side in Eq.~(\ref{eq:KLDIV-2}) does not depend on $q({\cdot})$, we just have to maximize the first term in the right hand side, which is known as the evidence lower bound (ELBO), to minimize the KL divergence.  The ELBO provides the lower bound of the evidence $\log{p(Y)}$ because the KL divergence is non-negative.  Therefore, this procedure realizes the sufficient fitting of the observed data at the same time.  Indeed, the maximization of the ELBO serves the best explanation of the reduced dimension $Q$ of the latent variable.  
\section{Problem formulation in finance}
\subsection{Factor model}
Arbitrage pricing theory~\cite{ross2013arbitrage} assumes that the $D$-days expected return of an asset $r_n{\in}\mathbb{R}^N$ is explained by the factor model as
\begin{equation}
r_n = {\alpha}_n + F{\beta}_n + {\epsilon},
\label{eq:FACTOR}
\end{equation}
where ${\alpha}_n{\in}\mathbb{R}^D$ is an excess return, ${\beta}_n{\in}\mathbb{R}^Q$ is weight coefficients, $F{\in}\mathbb{R}^{D{\times}Q}$ is a factor matrix, and ${\epsilon}{\in}\mathbb{R}^D$ is an error term with zero mean and a finite covariance.  The factor model manifests that the return of the asset is originated from the returns of $Q$-factors.  In fact, without the excess return ${\alpha}_n$, the expected return of the factor model is derived as follows:
\begin{equation}
\mathbb{E}[r_n] = \mathbb{E}[F]{\beta}_n.
\label{eq:EXFACTOR}
\end{equation}
The special case of this formula with only one factor is known as the model of the capital asset pricing model, which is a cornerstone of the modern finance theory~\cite{harvey2016and}. \par
%
The weight coefficients ${\beta}_n$ in the factor model in Eq.~(\ref{eq:FACTOR}) can be interpreted as latent variables which explain the return of the asset.  Based on this idea, we introduce a nonlinear factor model as
\begin{equation}
r_n = f({\beta}_n).
\label{eq:NLLVM}
\end{equation}
This model is regarded as a latent variable counterpart of nonlinear factor model~\cite{nakagawa2018deep}.  Here, we employ the Student's $t$-process as the model of nonlinear mapping $f:\mathbb{R}^Q{\to}\mathbb{R}^D$.  In other words, the nonlinear factor model in Eq.~(\ref{eq:NLLVM}) is given by the TPLVM.  The nonlinear correlation of the latent variable factors depends on the specific form of the covariance function of the TPLVM, and the predicted return of the asset can be inferred by the predicted distribution.  Furthermore, the nonlinear factor model can be interpreted as a dimension reduction model when $Q<D$.  Hence we can expect to obtain the essential lower dimensional variable which explains the dynamics of the return of the asset. \par
\subsection{Portfolio theory}
Markowitz established the modern portfolio theory on the mean-variance portfolio.  In this theory, a portfolio consists of multi assets classes such as stock, bond, currency and commodity with their optimal allocations based on both individual and entangled risk of assets.  \par
The mean-variance portfolio is designed by the constrained quadratic programming problem with respect to the objective function as
\begin{equation}
w^TKw - {\lambda}(\mathbb{E}[r] - {\mu}),
\label{eq:MVFUNC}
\end{equation}
where $w{\in}\mathbb{R}^D$ is the weight coefficients of the portfolio, $K{\in}\mathbb{R}^{D{\times}D}$ is the covariance matrix of the returns, ${\lambda}$ is a Lagrangian multiplier, $r$ is the return of the portfolio and ${\mu}$ is the expected return of the portfolio.  In practical use, the return of the portfolio is quite hard to be estimated, whereby, without the constraint condition of the expected return, the mean-variance portfolio is often replaced by the minimum-variance portfolio with empirically estimated covariance matrix.
\section{Experiment}
\label{sec:EXP}
In this section, we test the performance of the minimum-variance portfolio with the TPLVM by comparing with that with the GPLVM.  Before proceeding, we explain the experimental dataset of our performance test.  \par
As the experimental data, we use the following global stock market indices: S\&P 500 (US), S\&P/TSX 60 (Canada), FTSE 100 (UK), CAC 40 (France), DAX (Germany), IBEX 35 (Spain), FTSE MIB (Italy), AEX (the Netherlands), OMX 30 (Sweden), SMI (Switzerland), Nikkei 225 (Japan), HKHSI (Hong Kong), ASX 200 (Australia), KOSPI (Korea), OBX (Norway), MSCI (Singapore).  These stock indices are sampled every month between Jun 1998 to Jun 2019 from the Bloomberg's data platform.  The statistics of the return of the stock indices are shown in Table~\ref{table:IDXSTAT}.  In this table, mean (Mean), standard deviation (Std.), the ratio of mean and standard deviation (R/R), skewness (Skew) and kurtosis (Kurtosis) of returns of the stock indices are presented. \par
With the use of the historical returns of the stock indices, we construct the minimum-variance portfolios based on the GPLVM ($\mathrm{Port}_G$) and TPLVM ($\mathrm{Port}_t$).  The covariance matrix of each portfolio is estimated by the covariance function with 120 past samples.  As the kernel function, we utilize the exponential kernel defined as
\begin{equation}
k_{\mathrm{Exp}}(x,x')={\theta}_1\exp{(-{\theta}_2^{-2}||x-x'||)}
\label{eq:EXPKER}
\end{equation}
with ${\theta}_l\;(l=1,2)$ being hyper parameters.  For the sake of brevity, the dimension of the latent variables are fixed $Q=1$.  Under these conditions, we compare the performance of the $\mathrm{Port}_G$ and $\mathrm{Port}_t$ by its annualized return (Return), annualized risk as the standard deviation of return (Risk), risk/return (R/R) as return divided by risk.
\begin{align}
 &\mathbf{Return} = \frac{12}{T}\sum_{t=1}^T R_t^{P}  \\
 &\mathbf{Risk} = \sqrt{\frac{12}{T-1}\times(R_t^{P}-\mu^{P})^2}\\
 &\mathbf{R/R} = \mathbf{Return}/\mathbf{RISK}
\end{align}
Here, $R_t^{P}$ indicates GPLVM or TPLVM portfolio return at time $t$, and $\mu^{P}= (1/T) \sum_{t=1}^T R_t^{P}$ denotes the average return of the GPLVM or TPLVM portfolio.  \par
Table~\ref{table:RSLT} shows the performances of the portfolios by comparing annual return, risk and return-risk ratio.  The sample period is separated into anterior half period (Jun 2008 - Jun 2013) and posterior half period (Jul 2013 - Jun 2019).  Note that the anterior half period contains the global financial crisis 2007-2008.  As is seen in this table, the $\mathrm{Port}_t$ outperforms the $\mathrm{Port}_G$ in the both half periods.  In particular, the difference of the annual return in the anterior half period is larger than that in the posterior half period.  It is said that the market volatility during the global financial crisis intensively fluctuated whereby non-Gaussian nature clearly emerged in the global stock market.  In such situation, the TPLVM is a consistent model to describe the intermittent volatility fluctuations.  Thus, we can construct a robust portfolio by the TPLVM based minimum-variance portfolio.
\begin{table*}[]
\caption{Statistics of global market indices}
\label{table:IDXSTAT}
\begin{tabular}{crrrrrrrr}
\hline
\multicolumn{1}{|c|}{}     & \multicolumn{1}{c|}{US}      & \multicolumn{1}{c|}{Canada}      & \multicolumn{1}{c|}{UK}      & \multicolumn{1}{c|}{France}   & \multicolumn{1}{c|}{Germany}   & \multicolumn{1}{c|}{Spain}   & \multicolumn{1}{c|}{Italy}   & \multicolumn{1}{c|}{Netherlands} \\ \hline
\multicolumn{1}{|c|}{Mean [\%]} & \multicolumn{1}{r|}{6.00}  & \multicolumn{1}{r|}{5.41}      & \multicolumn{1}{r|}{2.39}  & \multicolumn{1}{r|}{4.08}   & \multicolumn{1}{r|}{6.87}    & \multicolumn{1}{r|}{3.20}  & \multicolumn{1}{r|}{1.35}  & \multicolumn{1}{r|}{2.96}      \\ \hline
\multicolumn{1}{|c|}{Std. [\%]}  & \multicolumn{1}{r|}{14.93} & \multicolumn{1}{r|}{14.92}     & \multicolumn{1}{r|}{13.62} & \multicolumn{1}{r|}{18.12}  & \multicolumn{1}{r|}{21.13}   & \multicolumn{1}{r|}{20.66} & \multicolumn{1}{r|}{21.71} & \multicolumn{1}{r|}{19.13}     \\ \hline
\multicolumn{1}{|c|}{R/R}    & \multicolumn{1}{r|}{0.40}    & \multicolumn{1}{r|}{0.36}        & \multicolumn{1}{r|}{0.18}    & \multicolumn{1}{r|}{0.23}     & \multicolumn{1}{r|}{0.33}      & \multicolumn{1}{r|}{0.15}    & \multicolumn{1}{r|}{0.06}    & \multicolumn{1}{r|}{0.15}        \\ \hline
\multicolumn{1}{|c|}{Skew}     & \multicolumn{1}{r|}{-0.66}   & \multicolumn{1}{r|}{-0.92}       & \multicolumn{1}{r|}{-0.55}   & \multicolumn{1}{r|}{-0.38}    & \multicolumn{1}{r|}{-0.50}     & \multicolumn{1}{r|}{-0.17}   & \multicolumn{1}{r|}{0.03}    & \multicolumn{1}{r|}{-0.74}       \\ \hline
\multicolumn{1}{|c|}{Kurtosis}     & \multicolumn{1}{r|}{5.23}    & \multicolumn{1}{r|}{7.36}        & \multicolumn{1}{r|}{4.53}    & \multicolumn{1}{r|}{4.52}     & \multicolumn{1}{r|}{6.12}      & \multicolumn{1}{r|}{4.96}    & \multicolumn{1}{r|}{4.80}    & \multicolumn{1}{r|}{5.88}        \\ \hline
\multicolumn{1}{l}{}         & \multicolumn{1}{l}{}         & \multicolumn{1}{l}{}             & \multicolumn{1}{l}{}         & \multicolumn{1}{l}{}          & \multicolumn{1}{l}{}           & \multicolumn{1}{l}{}         & \multicolumn{1}{l}{}         & \multicolumn{1}{l}{}             \\ \hline
\multicolumn{1}{|c|}{}     & \multicolumn{1}{c|}{Sweden}  & \multicolumn{1}{c|}{Switzerland} & \multicolumn{1}{c|}{Japan}   & \multicolumn{1}{c|}{HongKong} & \multicolumn{1}{c|}{Australia} & \multicolumn{1}{c|}{Korea}   & \multicolumn{1}{c|}{Norway}  & \multicolumn{1}{c|}{Singapore}   \\ \hline
\multicolumn{1}{|c|}{Mean [\%]} & \multicolumn{1}{r|}{6.32}  & \multicolumn{1}{r|}{2.80}      & \multicolumn{1}{r|}{3.35}  & \multicolumn{1}{r|}{7.27}   & \multicolumn{1}{r|}{4.70}    & \multicolumn{1}{r|}{12.98} & \multicolumn{1}{r|}{10.72} & \multicolumn{1}{r|}{5.05}      \\ \hline
\multicolumn{1}{|c|}{Std. [\%]}  & \multicolumn{1}{r|}{19.51} & \multicolumn{1}{r|}{14.68}     & \multicolumn{1}{r|}{19.24} & \multicolumn{1}{r|}{23.46}  & \multicolumn{1}{r|}{12.40}   & \multicolumn{1}{r|}{28.80} & \multicolumn{1}{r|}{21.49} & \multicolumn{1}{r|}{21.71}     \\ \hline
\multicolumn{1}{|c|}{R/R}    & \multicolumn{1}{r|}{0.32}    & \multicolumn{1}{r|}{0.19}        & \multicolumn{1}{r|}{0.17}    & \multicolumn{1}{r|}{0.31}     & \multicolumn{1}{r|}{0.38}      & \multicolumn{1}{r|}{0.45}    & \multicolumn{1}{r|}{0.50}    & \multicolumn{1}{r|}{0.23}        \\ \hline
\multicolumn{1}{|c|}{Skew}     & \multicolumn{1}{r|}{-0.19}   & \multicolumn{1}{r|}{-0.73}       & \multicolumn{1}{r|}{-0.54}   & \multicolumn{1}{r|}{0.28}     & \multicolumn{1}{r|}{-0.69}     & \multicolumn{1}{r|}{1.39}    & \multicolumn{1}{r|}{-0.93}   & \multicolumn{1}{r|}{-0.26}       \\ \hline
\multicolumn{1}{|c|}{Kurtosis}     & \multicolumn{1}{r|}{5.29}    & \multicolumn{1}{r|}{6.11}        & \multicolumn{1}{r|}{4.75}    & \multicolumn{1}{r|}{5.78}     & \multicolumn{1}{r|}{4.54}      & \multicolumn{1}{r|}{11.63}   & \multicolumn{1}{r|}{6.84}    & \multicolumn{1}{r|}{6.81}        \\ \hline
\end{tabular}
\end{table*}
%
\begin{table}[]
\centering
\caption{Performance of $\mathrm{Port}_G$ and $\mathrm{Port}_t$}
\label{table:RSLT}
\begin{tabular}{|c|r|r|r|}
\hline
       & \multicolumn{1}{c|}{$\mathrm{Port}_G$} & \multicolumn{1}{c|}{$\mathrm{Port}_t$} & \multicolumn{1}{c|}{Difference} \\ \hline
\multicolumn{4}{|c|}{Anterior half (Jun 2008 - Jun 2013)}                                                \\ \hline
Return & -4.89\%                      & -2.63\%                    & 2.25\%                 \\ \hline
Risk  & 19.57\%                      & 18.33\%                    & -1.24\%                \\ \hline
R/R  & -0.25                        & -0.14                      & 0.11                   \\ \hline
\multicolumn{4}{|c|}{Posterior half (Jul 2013 - Jun 2019)}                                                \\ \hline
Return & 6.08\%                       & 6.30\%                     & 0.22\%                 \\ \hline
Risk  & 11.16\%                      & 10.56\%                    & -0.60\%                \\ \hline
R/R  & 0.54                         & 0.60                       & 0.05                   \\ \hline
\multicolumn{4}{|c|}{Whole period (Jun 2008 - Jun 2019)}                                               \\ \hline
Return & 0.64\%                       & 1.87\%                     & 1.23\%                 \\ \hline
Risk  & 15.92\%                      & 14.93\%                    & -0.99\%                \\ \hline
R/R  & 0.04                         & 0.12                       & 0.09                   \\ \hline
\end{tabular}
\end{table}
\section{Conclusion}
\label{sec:CNCL}
In the literature of Bayesian machine learning, the Gaussian process has been developed and utilized to the diverse area including finance.  It is. however, well known that the historical financial data follows non-Gaussian distributions.  The Student's $t$-process is proposed,  as the generalization of the Gaussian process, to model the observed data following the non-Gaussian distributions with fat-tails.  \par
In this paper, we proposed the TPLVM by incorporating the latent variables into the Student's $t$-process.  The TPLVM can be used to reduce the number of explanation variable following the non-Gaussian distributions with fat-tails.  The nonlinear correlation of the TPLVM is modelled by prescribed kernel functions.  The hyperparameters of the TPLVM can be determined by the method of maximum-likelihood.  As a robust parameter optimization, we presented the method of variational inference of the TPLVM, which utilize the information of prior distribution of latent variables.   \par
The problem of the portfolio optimization has been studied in both academia and industry.  We applied the TPLVM into the portfolio optimization with the use of the minimum-variance portfolio.  To test the performance of the proposed portfolio, we implemented the empirical analysis for the global stock market data and compared the $\mathrm{Port}_G$ with $\mathrm{Port}_t$.  It was shown that the $\mathrm{Port}_t$ outperforms the $\mathrm{Port}_G$ in the whole test periods because $\mathrm{Port}_t$ can capture the non-Gaussian nature of the global stock market especially in the period of the global financial crisis.  \par
The TPLVM can be applied other risk-based portfolios such as risk parity~\cite{qian2011risk}, maximum risk diversification~\cite{choueifaty2008toward}, and complex valued risk diversification~\cite{uchiyama2019complex}.  These applications are expected to show high-performance compared with conventional ones.  In addition, the TPLVM can be modified to a latent variable dynamical model to catch the nature of historical volatility fluctuations.  These ways of research are our future works.


\end{document}